# Extinction by plasmonic nanoparticles in dispersive and dissipative media

SHANGYU ZHANG,[1,2] JIAN DONG,[1,2] WENJIE ZHANG,[1,2,*] MINGGANG LUO,[3] LINHUA LIU[1,2,*]

[1]*School of Energy and Power Engineering, Shandong University, Jinan 250061, China*
[2]*Optics & Thermal Radiation Research Center, Shandong University, Qingdao 266237, China*
[3]*School of Energy Science and Engineering, Harbin Institute of Technology, 92 West Street, Harbin 150001, China*
*Corresponding author: liulinhua@sdu.edu.cn*

**Extinction of small metallic spheres has been well understood through the classical Mie theory when the host medium is dispersive and transparent. However, the role of host dissipation on the particulate extinction remains a competition between the enhancing and reducing effects on the localized surface plasmonic resonance (LSPR). Here, using a generalized Mie theory, we elaborate on the specific influence mechanisms of host dissipation on the extinction efficiency factors of a plasmonic nanosphere. To this end, we isolate the dissipative effects by comparing the dispersive and dissipative host with its transparent counterpart. As a result, we identify the damping effects of host dissipation on the LSPR including the resonance widening and amplitude reducing. The resonance positions are shifted by host dissipation, which cannot be predicted by the classical Fröhlich condition. Finally, we demonstrate that a wide-band extinction enhancement due to host dissipation can be realized away from the positions of LSPR.**

Plasmonic nanoparticles have been attracting a lot of research interests due to their ability for excitation of localized surface plasmonic resonance (LSPR) which is a result of the interactions between electromagnetic waves and plasmon oscillations [1–3]. LSPR excites strong light-matter interactions and hence is of fundamental importance in many fields, including but not limited to thermoplasmonics [4], solar light harvesting [5], photothermal therapy [6], biosensing and environment monitoring [7], and plasmonic nanolasers [8]. It is well known that both the wavelength and amplitude of LSPR are sensitive to environment (host) media [9–11], by the mechanism of the variant optical constants for the different media. However, to simplify the analytical model, the environment (host) medium is usually assumed to be electromagnetically transparent, which is also termed as the dielectric environment [12].

The propagation of electromagnetic waves in the host medium is determined by means of the complex refractive indices $n$ depending on the wavelength $\lambda$ [i.e. $n(\lambda) = n' + in''$]. Dispersion $n'(\lambda)$ is necessarily connected to dissipation $n_2''(\lambda)$ through the Kramers-Kronig relations [13]. Strictly speaking, vacuum is the only nondispersive and transparent medium. Real materials like dielectrics are still regarded as nondispersive and electromagnetically transparent media, or even as dispersive transparent media for some wavelength bands over which the normal dispersion dominates and the absorption losses are negligible. However, dissipation cannot always be safely ignored as in the anomalous or highly dispersive regions, and hence, the couple effects of dispersion and dissipation of the host media on light scattering by particles must be carefully considered to avoid misleading outcomes.

In this Letter, we study the extinction of the plasmonic nanospheres embedded in a truly dispersive and dissipative medium and compare with that in its simplified dispersive and transparent counterpart for the two reasons:

(i) It is imperative to quantify the deviation of the optical extinction cross sections between a real material and its simplified transparent counterpart. The real host media surrounding the metal particles are both dispersive and dissipative in many applications, for example, tissues in the photothermal therapy [14], active layers in the polymer solar cells [15], and aqueous solution in the radio frequency heating [16]. However, the traditional scattering theory neglects the host dissipation [1–3,17], which has been used in many cases until now.

(ii) It helps to study the impacts of host dissipation on the LSPR. Although dissipation of host has been integrated into the generalized scattering theory in these years [18,19], the study of the dissipative effects on the LSPR has suffered a controversy. It is argued that the wavelength-independent constant dissipation (Kramers-Kronig inconsistent) will reduce and widen the LSPR [20,21], whereas a real dissipative material [like Poly(3-

hexylthiophene), P3HT] may enhance and sharpen the LSPR compared with another transparent host [like Poly(methyl methacrylate), PMMA] [22–24].

It is worth noting, first, that the expansion theory of frequency-domain electromagnetic scattering by a homogenous isotropic spherical particle is suitable for the dissipative host [25]. In other words, the expansion forms and the expansion coefficients of Mie theory are invariant no matter whether the dissipation of host is introduced or not. The expansion coefficients of scattered fields are [1]

$$a_l = \frac{\psi_l(x)\psi_l'(mx) - m\psi_l'(x)\psi_l(mx)}{\xi_l(x)\psi_l'(mx) - m\xi_l'(x)\psi_l(mx)}, \quad (1)$$

$$b_l = \frac{\psi_n'(x)\psi_l(mx) - m\psi_l(x)\psi_l'(mx)}{\xi_l'(x)\psi_l(mx) - m\xi_l(x)\psi_l'(mx)}, \quad (2)$$

where $\psi_l$ and $\xi_l$ are the Riccati-Bessel functions and the prime here indicates differentiation with the argument. Moreover, the relative refractive index $m$ and the size parameter $x$ are

$$m = \frac{n_1}{n_2}, \quad x = n_2 x_0 = n_2 k_0 a, \quad (3)$$

where $n_1$ and $n_2$ are, respectively, the complex refractive indices of sphere and host, $x_0$ is the vacuum size parameter, $k_0 = 2\pi/\lambda$ is the vacuum wave number with $\lambda$ the vacuum wavelength, and $a$ is the radius of the sphere.

In the context of the far-field scattering, however, considering the dissipative host compounds the situation due to the facts that the derived optical cross sections are dependent on the distance from the particle and, thus, cannot characterize the particle itself [18]. To obtain the distance-independent properties of the particle, the researchers adopt an actual experimental configuration, which models the extinction as the difference between the reading of a forward detector in the presence of the particle and that in the absence [18,26,27]. As a result, the generalized extinction efficiency considering the dissipation of host can be expressed as [27]

$$Q_{\text{ext}} = \frac{2\pi}{x'} \operatorname{Re} \sum_{l=1}^{\infty} \frac{1}{x}(2l+1)(a_l + b_l), \quad (4)$$

where $x'$ is the real part of size parameter $x$.

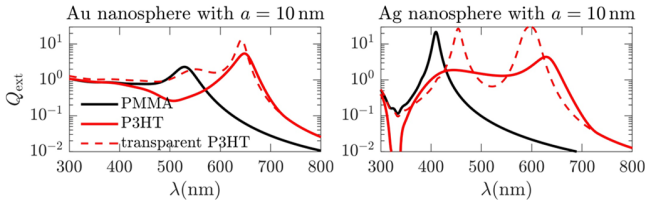

Fig. 1. Extinction efficiencies $Q_{\text{ext}}$ of Au (left) and Ag (right) nanospheres versus wavelength $\lambda$ in PMMA, P3HT, and transparent P3HT. It is important to note that, in the frequency band of interest, PMMA and P3HT are transparent and dissipative, respectively. The radius of the nanospheres is $a = 10$ nm.

Figure 1 plots $Q_{\text{ext}}$ of Au and Ag nanospheres with the radius of 10 nm in PMMA, real and dissipative P3HT, and transparent P3HT. It is worth noting that PMMA and P3HT are the transparent and dissipative media, respectively, in the wavelength region of 300-800 nm. The optical constants of the above materials are obtained from Refs. [28] and [23]. Our results have been verified to be consistent with the results calculated from the open source code recently provided by Mishchenko et al [29].

We first focus on the lines of Au in PMMA and truly dissipative P3HT in Fig. 1(a), which have been studied by Khlebtsov [23, Fig. 14] and Peck et al. [22,24]. Peck et al. [22,24] have concluded that plasmonic resonance in dissipative P3HT is stronger and narrower than that in dielectric PMMA host, which can be indeed seen in Fig. 1(a). They have attributed this narrowing mechanism to the rapid-wavelength-varying dissipation of P3HT. However, the narrowing mechanism has two flaws that inspires us to penetrate the effects of host dissipation on the plasmonic resonances:

(a) The narrowing mechanism is distinctly counter to conventional cognization that the additional losses degrade the resonances.

(b) The narrowing mechanism fails for Ag nanospheres as depicted in Fig. 1(b), where the resonance in dissipative P3HT is much damped and widened than that in dielectric PMMA.

We note that replacing P3HT with PMMA will induce the simultaneous changes of both dispersion and dissipation, i.e. the changes of wavelength dependent $n_2'$ and $n_2''$. The change of two variables at the same time makes the problem sophisticated. As inspired by the study of optical chirality in dispersive and lossy media in Ref. [30], we artificially separate the effects of dispersion and dissipation on the extinction efficiencies by comparing a truly dispersive and dissipative host with its dispersive and transparent counterpart.

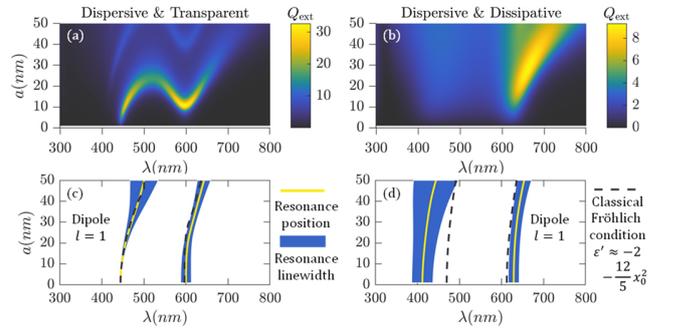

Fig. 2. Normalized extinction $Q_{\text{ext}}$ of Ag nanospheres versus wavelength $\lambda$ and radius of the spheres $a$ embedded in (a) transparent P3HT and (b) real P3HT. Electric dipole resonance positions and linewidths of Ag nanospheres are given in (c) transparent P3HT and (d) real P3HT. The positions are solved with Eq. (7) in the complex wavelength region (yellow solid line) and in real wavelength region (i.e. classical Fröhlich condition, dark dashed line). The linewidths solved in the complex wavelength region are illustrated with the blue area.

Figure 1 also gives the extinction efficiencies in transparent and dispersive P3HT, which helps to further study the dissipative effects on the resonance. As a result, the comparison between dielectric PMMA and real P3HT involves two successive processes which are the comparison between dielectric PMMA and transparent P3HT and the comparison between transparent P3HT and dissipative P3HT. According to the first comparison process, it is concluded from Fig. 1(a) that the dispersion of transparent P3HT, on the one hand, shifts the resonance of Au in PMMA from 520 nm to 640 nm, and on the other hand enhances the resonance amplitude and narrows the resonance width. However, in the second comparison process, we can observe that host dissipation widens and weakens the resonance, which agrees with the physical intuition that "adding losses to a plasmonic system will broaden plasmonic resonances". Besides, host dissipation induces a slight shift of the resonance position.

The comparison between P3HT and its transparent counterpart makes a focus on the influence of the host dissipation on the resonance. To show this comparison clearly, the extinction efficiencies of Ag nanoparticles versus the wavelength $\lambda$ and the radius of particles $a$ embedded in transparent P3HT and real P3HT are plotted in Figs. 2(a) and (b), respectively. As seen from Figs. 2(a) and (b), we can conclude that dissipation of P3HT will not only damp the resonance including the resonance widened and amplitude reduced but also shift the resonance position.

In the following, the resonance conditions and the resonance positions in dissipative host are studied. We consider the electric dipole term here as [31]

$$a_1 = \frac{-i\frac{2}{3}(\varepsilon-1)x^3}{\varepsilon+2-\frac{3}{5}(\varepsilon-2)x^2-i\frac{2}{3}(\varepsilon-1)x^3}, \quad (5)$$

where $\varepsilon = \varepsilon_1/\varepsilon_2$ is the relative dielectric function between the sphere and the host. The dipole resonance happens at the zero dominator of $a_1$, and hence, the resonance condition can be written in detail as

$$\varepsilon = -2 \times \frac{1+\frac{3}{5}x_0^2+\frac{i}{3}n_2x_0^3}{1-\frac{3}{5}x_0^2-\frac{2i}{3}n_2x_0^3}$$

$$\approx -2\left(1+\frac{6}{5}x_0^2+in_2x_0^3+\frac{18}{25}x_0^4+\frac{7i}{5}n_2x_0^5\right). \quad (6)$$

Considering the two orders of $x_0$, we obtain the resonance condition as

$$\varepsilon \approx -2 - \frac{12}{5}x_0^2. \quad (7)$$

There are two ways to solve resonance position according to the above equation. The first is solving the point in the real wavelength region which satisfies the classical Fröhlich condition of $\varepsilon' = -2 - 12x_0^2/5$, i.e. the real part of Eq. (7). The second way is to solve Eq. (7) in the complex wavelength or frequency domain [25,32]. Intriguingly, it has been proved by Fuchs and Kliewer [32] that the real part of solution represents the resonance position while the imaginary part represents the resonance half-width. The complex frequency method has been effectively used for the modal analysis of plasmonic nanoresonators [33] and for the resonant poles extraction for a dielectric sphere [34,35].

Figures 2(c) and (d), which correspond to the cases in Fig. 2(a) and (b), respectively, show the dipole resonance positions and linewidths in the two above ways. It is well known that the two ways give the close resonance positions when the host is transparent. This can be verified from Fig. 2(c) since the resonance positions solved by the classical Fröhlich condition are close to those from the complex-wavelength solutions. However, when the host dissipation is considered in Fig. 2(d), the resonance position described by the classical Fröhlich condition is away from that solved by the complex-wavelength-domain solutions. The difference of resonance positions between two ways is well beyond the resonance linewidth, which means that the classical Fröhlich condition gives inaccurate resonance positions when the host media are dissipative.

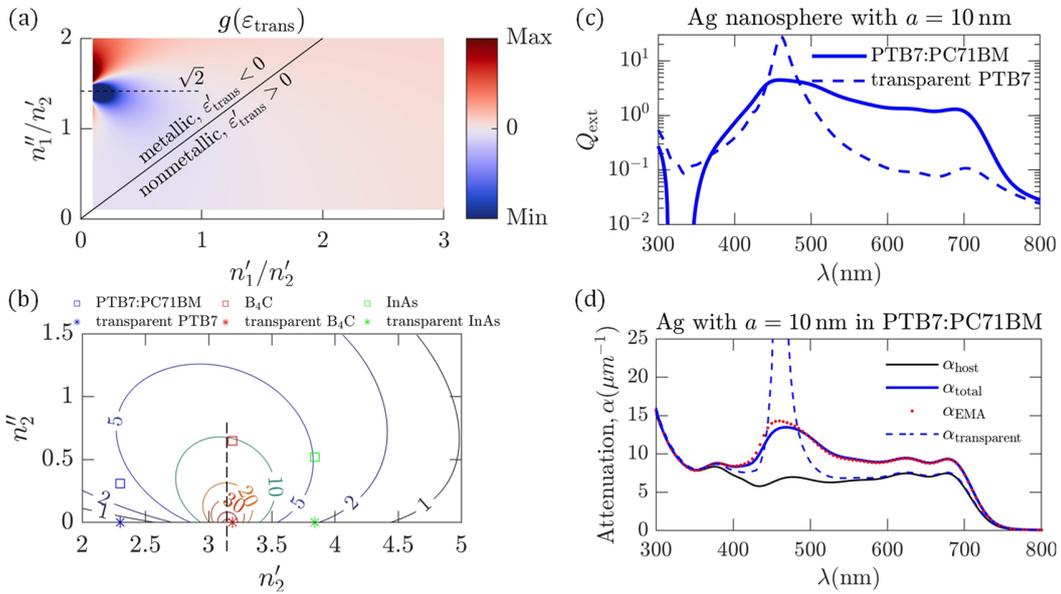

Fig. 3. (a) $g(\varepsilon_{\text{trans}})$ in Eqs. (11) and (12) versus $n_1'/n_2'$ and $n_1''/n_2'$. (b) Normalized extinction of Ag particles with the vacuum size parameter $x_0 = 0.1$ versus $n_2'$ and $n_2''$. The complex refraction index of Ag is taken at the wavelength of 770 nm. (c) Normalized extinction of Ag with radius of 10 nm in truly dissipative PTB7:PC71BM and its transparent counterpart. (d) Attenuation coefficients of Ag in PTB7:PC71BM host. The used volume fraction of Ag nanospheres in calculations is $f = 0.02$. $\alpha_{\text{host}}$ represents the attenuation coefficients arose from the host, $\alpha_{\text{total}} = \alpha_{\text{host}} + 3fQ_{\text{ext}}/4a$ is the total attenuation of the host and the Ag extinction considering the host dissipation, $\alpha_{\text{transparent}} = \alpha_{\text{host}} + 3fQ_{\text{ext}}^{\text{trans}}/4a$ means the total attenuation from the host and the Ag extinction neglecting the host dissipation, and $\alpha_{\text{EMA}}$ is calculated according to the effective medium approximation [36].

The above damping mechanism is discussed for the specific case. To obtain some general results, we note that the electric dipole term $a_1$ of Eq. (5) is the function of $\varepsilon$ and $x$. We can write the relative dielectric function $\varepsilon$ as

$$\varepsilon = \frac{\varepsilon_{\text{trans}}}{(1+i\tau)^2}, \tag{8}$$

where

$$\varepsilon_{\text{trans}} = \left(\frac{n_1'}{n_2'} + i\frac{n_1''}{n_2'}\right)^2, \tag{9}$$

is the relative dielectric function between the particle and transparent host and

$$\tau = n_2''/n_2', \tag{10}$$

represents the magnitude of the host dissipation. In this regard, we can expand the extinction contributed by the electric dipole at $\tau \approx 0$ with

$$Q_{\text{ext}} = Q_{\text{ext}}^{\text{trans}} + \tau \times 4x' g(\varepsilon_{\text{trans}}) + O(\tau^2), \tag{11}$$

where $Q_{\text{ext}}^{\text{trans}}$ is the normalized extinction for the transparent counterpart of host and

$$g(\varepsilon_{\text{trans}}) = \text{Im}\left[\frac{-4i(\varepsilon_{\text{trans}}-1) + 2i(\varepsilon_{\text{trans}}^2 - 4)}{(\varepsilon_{\text{trans}}+2)^2}\right]. \tag{12}$$

Therefore, the magnitude of $g(\varepsilon_{\text{trans}})$ controls the deviation of the real extinction with the transparent extinction $Q_{\text{ext}}^{\text{trans}}$, while the sign of $g(\varepsilon_{\text{trans}})$ determines if the absorbing host enhances or diminish the extinction. We plot $g(\varepsilon_{\text{trans}})$ in the space of $n_1'/n_2'$ and $n_1''/n_2'$ in Fig. 3(a). It is important to note that $g(\varepsilon_{\text{trans}})$ is irrelevant to the dissipation of host and relates the properties of host with $n_2'$. Therefore, as described above, the resonance condition in transparent host can be represented by the classical Fröhlich condition of $n_1'/n_2' \approx 0$ and $n_1''/n_2' \approx \sqrt{2}$ which is given by the dashed line in Fig. 3(a). It is concluded that the plasmonic resonance in transparent host is always diminished by the dissipation of the host, since $g(\varepsilon_{\text{trans}}) < 0$ near the resonance points in the transparent host.

However, an intriguing phenomenon occurs in Fig. 3(a), which is the enhanced extinction due to the host dissipation. The enhancement happens in the red region where the pairs of $n_1'/n_2'$ and $n_1''/n_2'$ are away from the resonance points in transparent host. In the following, we adopt a specific case to illustrate this extinction enhancement due to the host dissipation. Fig 3(b) illustrates the normalized extinction of an Ag particle with the vacuum size parameter $x_0 = 0.1$. The complex refraction indices of the Ag particle are adopted at the wavelength of 770 nm as $n_1 = 0.07 + 4.67i$. The extinction results are plotted in the space of the refractive indices of the host, $n_2'$ and $n_2''$, to clearly observe the effects of host dissipation $n_2''$. Therefore, we can again observe that, if the plasmonic resonance is met, the introduction of any dissipation of host will diminish the resonance. However, away from the resonance, the introduction of the $n_2''$ can indeed enhance the extinction. In Fig. 3(b), we also give two real materials of which the dissipation largely enhances the extinction of Ag particles, PTB7:PC71BM and InAs.

Finally, to illustrate the effects of host dissipation on the plasmonic particle with a real example, we plot the normalized extinction of Ag nanospheres in PTB7:PC71BM from 300 nm to 800 nm in Fig. 3(c). Comparing the extinction of Ag nanosphere in PTB7:PC71BM with its transparent counterpart, we concluded that the host dissipation produces a long-wavelength-range enhancement from about 500 nm to 700 nm. In what follows, we study the effects of the extinction enhancement on the attenuation coefficients. Given the volume fraction of Ag nanospheres $f = 0.02$, Fig. 3(d) shows the attenuation coefficients of the only host $\alpha_{\text{host}}$, the total attenuation arose from both extinction of Ag and the host $\alpha_{\text{total}} = \alpha_{\text{host}} + 3fQ_{\text{ext}}/4a$, the total attenuation by the extinction of Ag in transparent host and the host $\alpha_{\text{transparent}} = \alpha_{\text{host}} + 3fQ_{\text{ext}}^{\text{trans}}/4a$, and the attenuation calculated from effective medium approximation $\alpha_{\text{EMA}}$ according to the equation in Ref. [36]. We can conclude that the enhanced extinction will further enhance the attenuation coefficients, which can be verified by the results of effective medium approximation.

In summary, we study the extinction by plasmonic nanoparticles embedded in a dispersive and dissipative medium in this Letter. We reexamine the former and counterintuitive conclusion that the localized surface plasmonic resonance (LSPR) can be enhanced by the dissipation of environment (host) media. By comparing the dispersive and dissipative media with its transparent counterpart, we identify the damping of the LSPR due to the host dissipation which is consistent with conventional wisdom, including the resonance widening and amplitude reducing. Besides, we observe the resonance shifts induced by the host dissipation. Giving the general resonance conditions, we illustrate the misleading prediction of resonance positions by the classical Fröhlich condition. Furthermore, away from the resonance, we find that the host dissipation can indeed enhance the extinction and further enhance the attenuation coefficients. In view of the growing interests in light-matter interactions, we expect that our results will aid the optimal design of plasmonic nanoparticles embedding the dissipative media, especially in the context of the enhanced extinction.

**Funding.** National Natural Science Foundation of China (52076123); China Postdoctoral Science Foundation (2021M691906, 2021M700991); Shandong Provincial Natural Science Foundation (ZR2020QE194).

**Disclosures.** The authors declare no conflicts of interest.